# Hexagonal Boron Nitride Phononic Crystal Waveguides

Yanan Wang,[†,‡] Jaesung Lee,[†,‡] Xu-Qian Zheng,[†,‡] Yong Xie,[†,∥] and Philip X.-L. Feng[*,†,‡]

[†]*Department of Electrical Engineering & Computer Science, Case School of Engineering, Case Western Reserve University, Cleveland, Ohio 44106, United States*

[‡]*Department of Electrical & Computer Engineering, Herbert Wertheim College of Engineering, University of Florida, Gainesville, Florida 32611, United States*

[∥]*School of Advanced Materials & Nanotechnology, Xidian University, Xi'an, Shaanxi 710071, China*

[*]Email: philip.feng@ufl.edu

## Abstract

Hexagonal boron nitride (h-BN), one of the hallmark van der Waals (vdW) layered crystals with an ensemble of attractive physical properties, is playing increasingly important roles in exploring two-dimensional (2D) electronics, photonics, mechanics, and emerging quantum engineering. Here, we report on the demonstration of h-BN phononic crystal waveguides with designed pass and stop bands in the radio frequency (RF) range and controllable wave propagation and transmission, by harnessing arrays of coupled h-BN nanomechanical resonators with engineerable coupling strength. Experimental measurements validate that these phononic crystal waveguides confine and support 15–24 megahertz (MHz) wave propagation over 1.2 millimeters. Analogous to solid-state atomic crystal lattices, phononic bandgaps and dispersive behaviors have been observed and systematically investigated in the h-BN phononic waveguides. Guiding and manipulating acoustic waves on such additively integratable h-BN platform may facilitate multiphysical coupling and information transduction, and open up new opportunities for coherent on-chip signal processing and communication via emerging h-BN photonic and phononic devices.

The considerable research efforts on exploring the unconventional properties of van der Waals (vdW) layered materials have already led to exciting breakthroughs across a variety of disciplines from fundamental science to device engineering [1,2,3,4,5,6]. Among families of vdW crystals, hexagonal boron nitride (h-BN), having prevailed as gate dielectric and passivation layers in two-dimensional (2D) electronics and optoelectronics [7], is emerging as an attractive material platform for nanophotonics and quantum optics [8,9,10]. The ultrawide bandgap (5.9 eV) of h-BN is beneficial not only for providing excellent electrical insulation, but also for hosting robust single photon emitters (SPEs) even at room temperature [10]. These defect-related quantum emitters exhibit large Debye-Waller (DW) factors and megahertz (MHz) photon count rates among the brightest SPEs reported so far [10,11]. The layered structure of h-BN also endorses new and unparalleled flexibility in additive, back-end-of-line, and hybrid device integration free from lattice matching constraints, making it an excellent candidate for implementing on-chip quantum information processing and sensing functions.

Development of future quantum circuitries will require on-chip integration of multiple physical components in a way that the combined advantages of the hybrid system mitigate the weaknesses of individual constituents [12]. In parallel with photonic components, such as photonic crystal cavities and optical resonators, which have been preliminarily demonstrated in h-BN crystals [13,14], phononic wave devices also play crucial roles in such hybrid schemes. Traveling at significantly slower speeds than the luminal speed of photons, phonons—the quanta of mechanical vibrations—have been suggested as better carriers allowing information to be stored, filtered and delayed over comparatively small length-scales but with high fidelity [15,16,17]. It has been widely demonstrated that mechanical motion can mediate energy/information transduction among different physical domains, as well as bridge the classical and quantum regimes. For instance, the conversion between electrical signal and mechanical vibration has been well established in micro/nanoelectromechanical systems (M/NEMS), which are the backbones of today's state-of-the-art commercial timing and inertial sensing devices [18,19], and the most sensitive probes in exploring fundamental science and limits of measurement [20,21,22]. Coherent photon-phonon interactions can be achieved via radiation pressure forces in optomechanical cavities [15,23,24]. In the quantum regime, mechanical resonators can be coupled to artificial atoms embodied as charge qubits or spin qubits through Coulomb interactions or magnetic dipole forces, respectively [12,16,25,26]. Specifically and intriguingly, spin-mechanical coupling schemes have been recently proposed and theoretically investigated based on the h-BN quantum emitters [27,28].

Thanks to the layered structure of its crystals, h-BN renders a unique combination of mechanical properties with very high in-plane stiffness and strength but low flexural rigidity (especially for monolayer and few-layer structures). It is predicted and validated as an excellent structural material, with a theoretical Young's modulus $E_Y \approx 780$ GPa and a breaking strain limit up to $\varepsilon \approx 20\%$ [29,30,31]. The fundamental mechanical properties have been explored in suspended h-BN structures [30,32]; however, only very few experimental demonstrations of mechanical devices have been reported to date [31,33,34]. To address the aforementioned needs and challenges for future integrated hybrid systems, it is desirable to explore new h-BN nanomechanical devices, especially ones capable of manipulating waves coherently.

In this work, we take an initiative to experimentally demonstrate one type of building blocks essential for future on-chip phononic integration, namely phononic crystal waveguides, in h-BN crystals. Distinct from individual nanomechanical resonators reported previously [31,34], phononic crystal waveguides comprise periodic arrays of coupled mechanical structures with architected unit cells (Figure 1). For individual nanomechanical resonators, the vibrational motion is restricted to the stationary eigenmodes. Whereas, vibration can propagate through coupled resonators, resulting in either constructive or destructive phonon transmission. Hence, by assembling coupled resonators into long-range periodic structures, we can create phononic crystal structures and consequential phononic bands similar to the atomic lattices in crystalline solids and photonic crystal lattices, and their associated band structures. Such devices not only enable fundamental exploration of lattice-based solid-state phenomena including dispersive relation, energy transport, nonlinear dynamics, and topological states in the phononic domain, but also facilitate device functionalities, such as on-chip routing and filtering of radio-frequency (RF) acoustic waves [35,36].

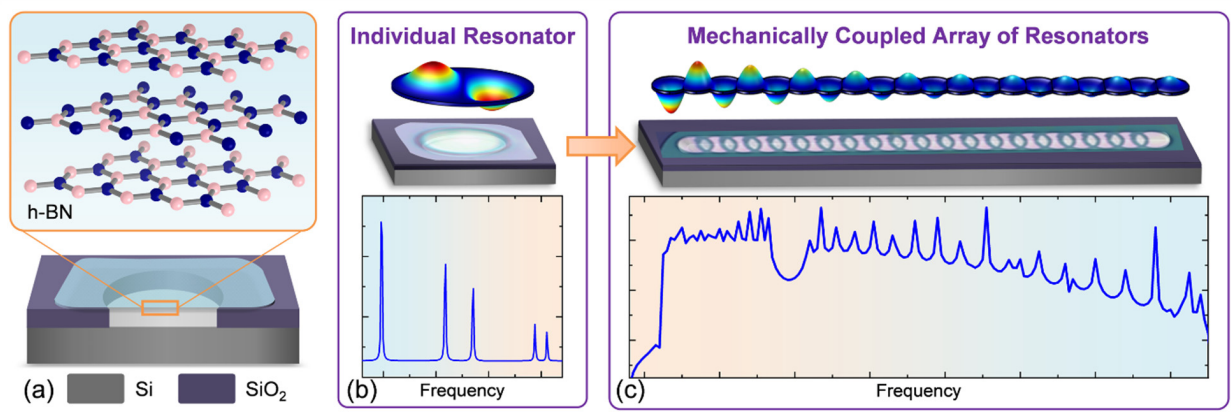

**Figure 1.** Evolution from an individual nanomechanical resonator to a coupled resonator array as a phononic waveguide. (a) Cross-sectional schematic of suspended h-BN drumhead resonator over patterned $SiO_2$/Si substrate. Typical mode shapes and frequency characteristics of (b) individual and (c) coupled h-BN resonators. Panels b and c illustrate the frequency responses evolving from discrete eigenmodes of the individual resonator to continuous transmission bands of the coupled resonators. Optical microscopy images of the devices are embedded in the schematic illustrations.

The basic concept in Figure 1 is enabled and reinforced by the agile, additive features in device nanofabrication in the h-BN platform, evading conventional lithographic patterning and its associated chemical resist and contamination. As illustrated in Figures 1 and S1, the periodic structures are defined on the commonly used oxidized silicon (290 nm $SiO_2$ on Si) supporting substrates, taking advantage of the well-established patterning and etching techniques with high spatial resolution (see Supporting Information). Then, a suite of specially engineered, completely dry exfoliation and stamp-transfer techniques are employed to create large-area suspended h-BN structures [31,33,37]. In as-fabricated devices, sizable elastic impedance mismatch is created between the suspended and supported regions of h-BN, thereby the acoustic energy can be efficiently confined within the suspended waveguides. The periodic elastic energy potential profile results in the formation of phononic band structure. This device fabrication approach can also help preserve the quantum signatures of emitters in h-BN free from any wet chemistry or surface contamination [38], favorable for developing future hybrid quantum devices.

Figure 2 presents the quantitative designs of the devices with modeling results. Quasi-1D chain of edge clamped circular resonators are designed to be overlapped with the neighboring ones, in order to achieve strong mechanical coupling. The effective unit cell can be simplified as a near oval shape highlighted in Figures 2a and S1. The period of the lattice is set to be $a$=8.25 μm and the width of the cell is $w$=12 μm, owing to the predicted frequency dispersion with prominent bandgap lying in the MHz range (Figures S2 and S3). According to the finite element method (FEM) simulations (COMSOL Multiphysics™), the thickness of h-BN layers is chosen to be larger than 50 nm to ensure the fabricated nanomechanical structures follow the plate model (detailed discussion in the Supporting Information) and immune to local strain inhomogeneity [39]. The experimental and simulation results presented in the Main Text are attained based on the aforementioned cell geometry settings with a thickness of h-BN being 120 nm, without additional notation. Additional devices with varied unit cell dimensions are showcased in the Supporting Information (Figures S1 and S6).

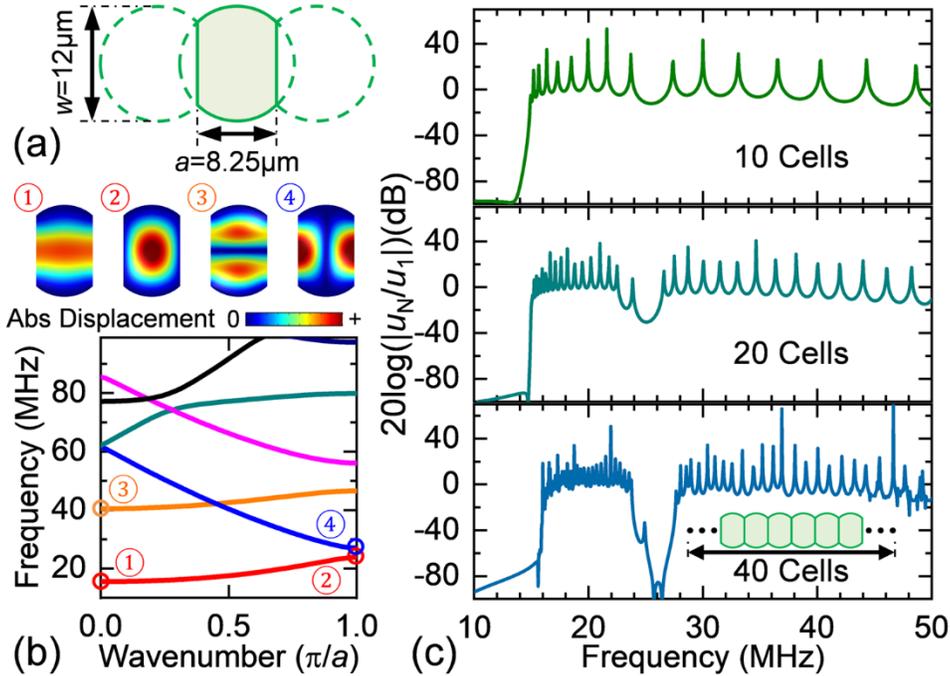

**Figure 2.** Phononic waveguide design. (a) Design of spatially overlapped circular h-BN nanomechanical structure and the effective unit cell of the quasi-1D phononic crystal lattice highlighted in light green shade. Without specific notation, the phononic crystal waveguides presented in the following figures have a lattice period of $a$=8.25 μm, a unit cell width of $w$=12 μm equal to the diameter of the circle. (b) Phonon dispersion curves numerically calculated for a 120 nm thick h-BN waveguide with the unit-cell mode shapes at different frequencies (①–④) labeled in the plot. (c) The transmission characteristics and performance of phononic waveguides computed with numbers of cells $N$ varying from 10 to 40 (from top to bottom).

The phononic band structure of the designed waveguide is revealed as the simulated phonon dispersion relation curves shown in Figure 2b, along with the mode shapes of the unit cell for the first and second lowest eigenfrequency branches at the wavenumber $k$=0 and $\pi/a$ points, respectively. The frequency spectrum is divided into multiple regions. The frequency range below the first phonon band is called stop band, in which no mechanical mode can be supported by the waveguide. Analogous to electron transport in atomic lattices of crystalline semiconductors, bandgaps form between transmission bands, within which phonon propagation is also suppressed. At higher frequencies, neighboring phonon bands start merging and mode crossing/anticrossing occur, related to the phenomenon of strong mode coupling which can result in the mode type transformation and energy exchange between different phonon branches.

It is worth mentioning that such band structure simulation is fulfilled with Floquet-Bloch periodicity conditions $u(x+a) = e^{ika}u(x)$, where $u(x)$ is the displacement at position $x$, $a$ is the lattice period and $k$ is the wavenumber ($0 \leq k \leq \pi/a$), assuming the number of cells as infinite. However, for device fabrication, we need to balance between the waveguide performance and the device footprint to determine a practical number of cells. Therefore, we also study the influence of the constituent cell numbers ($N$=10, 20, 40) numerically, as summarized in Figure 2c. The relative amplitude in frequency-domain simulations, or transmission $T_{N-1}$=20 log($|u_N/u_1|$) in dB, is defined as the ratio between the areal averaged displacement magnitude ($|u_N|$=$\int_{Area}|u|$dxdy/$A_{Cell}$) of the first and the last cells (note in analysis and simulation of individual cells, $u$=$u(x,y)$, no longer quasi-1D), with excitation force loaded at the first one. Reliant on the constructive or destructive interference conditions, the displacement amplitude of the last cell can exceed (>0 dB) or fall short (<0 dB) from the amplitude of the first cell. Agreeing with the band structure simulation, clear stop bands (<15 MHz) can be observed with displacement ratio lower than -90 dB for all the presented cases. Conversely, there is no noticeable relative amplitude decline at the predicted bandgap frequencies (24–27 MHz) for the lattice with $N$=10. For $N$=20 and 40 cases, attenuation of ~30 dB and ~90 dB can be achieved, respectively. Considering the difficulty in obtaining h-BN flakes with very large uniform areas, the number of cells is set around 20 for device fabrication.

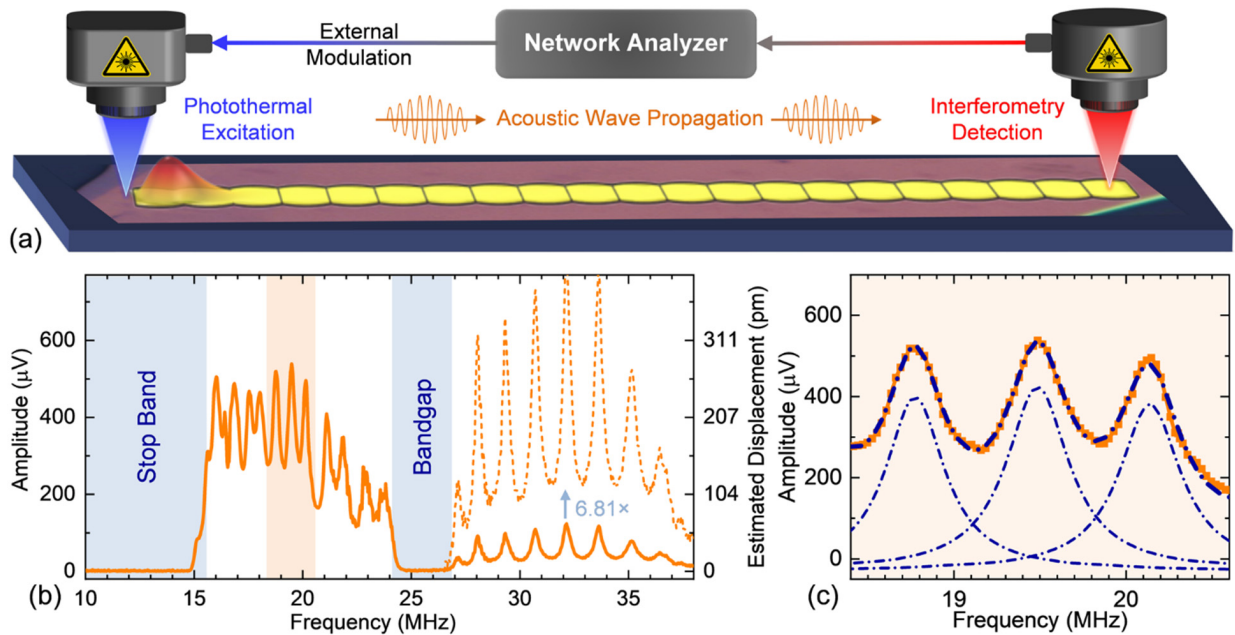

**Figure 3.** Frequency-domain characterization of h-BN phononic waveguide. (a) Measurement scheme of frequency-domain characterization. The waveguide is photothermally excited by an amplitude-modulated 405 nm laser at one end (near the leftmost cell in panel a) and the resultant phonon propagation is detected via a 633 nm laser interferometry system at the other end (rightmost cell in panel a). In real experiments, the 405 nm and 633 nm lasers are focused through the same 50× objective with tunable spatial separation up to ~200 μm. Optical microscopy image and simulated mode displacement profile at the excitation frequency $f$=14 MHz are embedded in the schematic illustration. (b) Measured frequency responses of the waveguide over a broad range, with the dashed line showing the response corrected according to position dependent detection efficiency ($|u_{antinode}/u_{node}|\approx6.81$) and right axis showing the estimated displacement. (c) Zoomed-in spectrum of the frequency range highlighted in light orange shade in panel b and each peak is fitted with finite-$Q$ harmonic resonance function.

To investigate the frequency-domain characteristics of the h-BN phononic waveguides experimentally, we employ a customized ultrasensitive laser interferometry system (Figures 3a and S4). The conventional vibration excitation techniques relying on electromechanical coupling are not readily applicable to the insulating h-BN crystals. Also, the piezoelectric effect in h-BN only becomes accessible when the flake is thinned down to few layers [40]. All-optical actuation and detection scheme is thus suitable for the characterization of these phononic devices, in which the suspended h-BN crystal and the patterned $SiO_2$/Si substrate form an interferometer. The system is equipped with an amplitude-modulated 405 nm blue laser as the driving beam and a continuous-wave 633 nm red laser as the detecting beam. Both beams are focused onto the devices through the same 50× long-working-distance objective, while the position of the focused laser spots can be individually controlled by a set of mirrors and beam splitters with spatial separation up to ~200 μm. Flexural motion of the suspended h-BN waveguide is photothermally actuated by the modulated 405 nm laser at one end. The acoustic waves with frequencies in the transmission bands can propagate through the waveguide and alter the interferometry condition of 633 nm laser at the other end of the waveguide. Hence, the mechanical displacement of the device can be transduced into the intensity variation of the reflected 633 nm laser beam and read out by a photodetector with frequency swept by a network analyzer.

Measured frequency-domain responses are shown in Figures 3 and 4 (and more data from additional devices in Supporting Information). As predicted, closely packed resonance modes develop into continuous transmission bands with well-defined phononic stop band and bandgap features. In terms of characteristic frequencies, the measured spectrum shows excellent agreement with the simulated one and the dispersive curves (Figure panels 4c–4f). The first phonon band spans from 15 to 24 MHz, and the second phonon band is from 27 to 40 MHz, separated by a prominent bandgap of ~3 MHz. The amplitude measured from the second phonon band is noticeably smaller than the one from the first band, because the detecting 633 nm laser spot is still parked at the center of the cell (no. 21), but this is now on the node lines of the unit-cell mode shapes ③ and ④ illustrated in Figure 2b. The lower branches are replotted in Figure S7 for the analysis and computation of the position-dependent displacement detection efficiency within any unit cell, to obtain the correction coefficient $|u_{antinode}/u_{node}| \approx 6.81$ for the second phonon band, based on a combination of measurements and simulations with practical laser spot size (1.5μm) on the device (see Supporting Information). The amplitude-corrected response of the second phonon band is plotted as a dashed line in Figure 3b, which corresponds to the actual displacement at the last cell (no. 21) in this band. The corrected response in the second band can be higher than that in the first band, consistent with simulation results in Figures 2c and 4d. The ratio of displacement amplitude between the bandgap and the first transmission band is approximately $10^{-2}$, corresponding to an attenuation of ~33.5 dB, also agreeing with the simulation (Figure panels 4d–4f).

Figure 4a depicts a system diagram that facilitates analyzing the signal transduction chain and gaining quantitative understanding of the intrinsic characteristics of the waveguide device. The measured raw data (amplitude plot in rms voltage, Figure 3, and transmission plot in dB, Figure 4e) include extrinsic responses and contributions from the peripheral transducers necessary for interfacing with and measuring the device, including the upstream photothermal actuation, the downstream motion detection modules, and their inner components (Figure 4a). Combining multi-stage signal transduction gain analysis and calibration measurement by replacing the waveguide device with a single h-BN drumhead resonator with dimensions similar to a unit cell and operating within the first phonon band (15–24MHz), we can obtain a relation that allows us to subtract the extrinsic contributions from the measured data (see Supporting Information). Upon application to the transmission data in Figure 4e, this yields the measured intrinsic transmission of the waveguide, shown in Figure 4f (without repeating the second band amplitude correction, dashed line as shown in Figure 3b), in very good quantitative agreement with the simulation in Figure 4d. In addition, with the analysis and calibration of displacement-to-voltage transduction in the motion detection module (Figure 4a), the corresponding displacement-domain data is also determined for the measured amplitude response shown in Figure 3b, exhibiting ~50 pm to 400 pm vibrations probed at cell 21 in the first and second phonon bands (refer to the right vertical axis of Figure 3b).

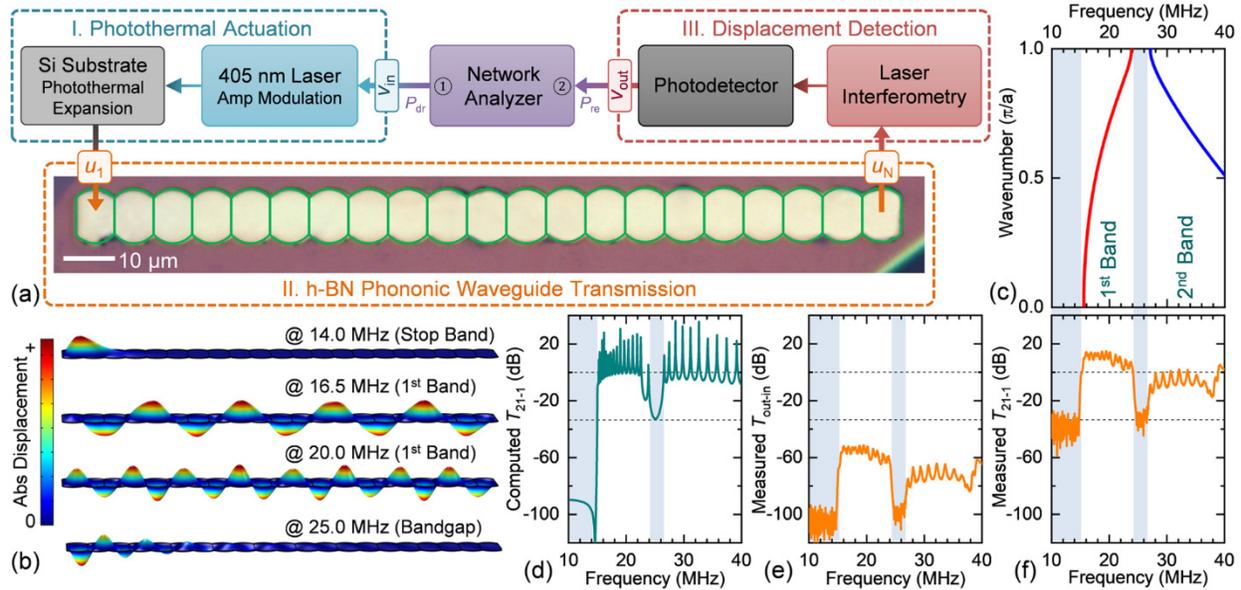

**Figure 4.** Signal transduction analysis and comparison between the measured and simulated frequency responses. (a) Signal transduction flowchart. $v_{in}$ and $v_{out}$ represent the input root-mean-square (rms) voltage to the system under test, from the network analyzer to the amplitude modulated 405 nm driving laser, and the output rms voltage of the photodetector fed into the network analyzer, respectively. $u_1$ and $u_N$ are the areal averaged displacement of the first and last cells, respectively. (b) Displacement profile of the h-BN waveguide at varied frequencies $f$=14, 16.5, 20, 25 MHz (from top to bottom). The excitation is applied near the leftmost cell, unit cell 1 in (a). (c) Numerical dispersion relation of the first and second lowest phonon branches around $k=\pi/a$ from eigenfrequency simulation. (d) Intrinsic transmission of the waveguide, $T_{21-I}=20\log(|u_{21}/u_1|)$, obtained from frequency-domain simulation. (e) Measured total transmission $T_{out-in}=20\log(|v_{out}/v_{in}|)$ plot converted from the measured amplitude versus frequency spectrum in Figure 3b. (f) Measured intrinsic transmission of the device, according to data in (e) and analysis and calibration from (a). The dashed lines in (d)–(f) delineate the 0 dB baseline and the 33.5 dB attenuation within the bandgap.

The phonon propagation behaviors can be directly visualized in the mode displacement analysis (Figure 4b). When one end of the waveguide is excited at a stop band frequency of 14 MHz, the mechanical motion is restricted to the excitation spot without propagation along the waveguide. When the excitation frequency escalates and reaches the first phonon band, propagating phonons with long wavelengths can be supported by the waveguide (@16.5 MHz in Figure 4b). As the frequency keeps rising, the wavelength of acoustic wave gradually decreases (@20 MHz in Figure 4b). When the half wavelength of the standing wave is in proximity to the same length scale of the unit cell, destructive interference among the reflected waves inside the waveguide leads to suppression of the mechanical motion. Therefore, at a bandgap frequency of 25 MHz, the excited vibration cannot travel through the waveguide (Figure 4b).

To further understand the phonon propagation dynamics in the h-BN phononic waveguide, temporal measurements are performed as well. To monitor the mechanical wave propagation in time domain, we build the time-domain measurement apparatus, as illustrated in Figures 5a and S10. The drive signal of the modulated 405 nm laser is shaped by a function generator. We introduce a single-frequency sinusoidal actuation signal with 15 cycles by enabling the RF burst mode of the function generator. The pulse train (in each burst period) gradually intensifies the mechanical motion of the device, and then the generated acoustic wave propagates through the phononic waveguide until it relaxes to below the thermal noise limit due to the dissipation during propagation. The displacement of the mechanical wave packet is probed at the end of the waveguide by the same 633 nm laser interferometry. But the electrical signal from the photodetector is filtered by a set of bandpass (BP) filters with 10–30 MHz bandwidth and then recorded using an oscilloscope. The results are averaged over 512 times by synchronizing the time-domain signals using the trigger from the function generator. The period of the burst signals is set as 10 ms to ensure the time interval between each excitation pulse train is long enough compared to the ring-down time of the mechanical motions so that the acoustic wave can propagate back and forth inside the waveguide for multiple cycles.

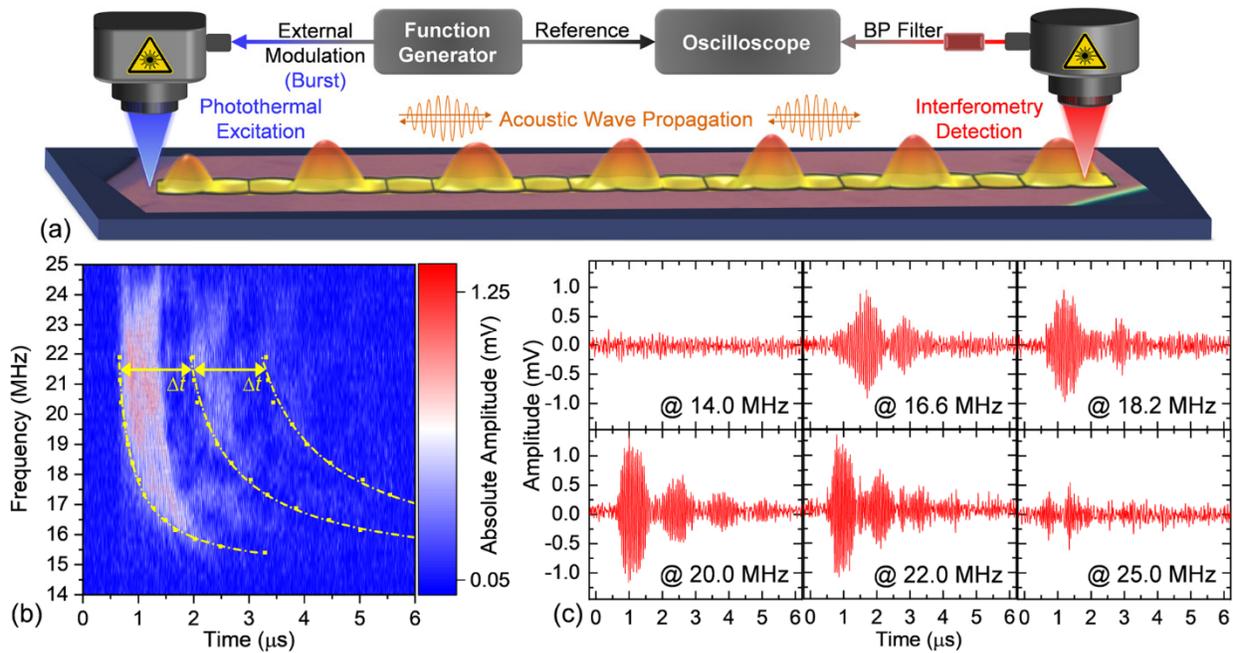

**Figure 5.** Temporal dynamics of phonon propagation. (a) Measurement scheme of time-domain characterization. Distinct from the frequency-domain measurement, the excitation laser is modulated by sinusoidal burst signals from a function generator (number of cycles *n*=15, burst period *p*=10 ms, pulse train shown in Figure S10a) and these burst signals are also synced to the oscilloscope as references. A set of bandpass (BP) filters are inserted in the detection circuit, which only allow the waves with frequencies of 10–30 MHz passing through. Optical microscopy image and simulated mode displacement at the excitation frequency *f*=18 MHz are embedded in the schematic illustration. (b) The color map represents the measured time-domain amplitude responses by sweeping the excitation frequency across the first phononic band. Δ*t* defines the time of the acoustic wave propagating a round trip along the waveguide. The yellow squares and dash lines mark the round-trip traveling time calculated from the eigenfrequency simulation and the frequency-domain simulation shown in Figure S11. (c) Selected traces of time-domain measurement from (b) at varied frequencies *f*=14, 16.6, 18.2, 20, 22, 25 MHz.

The measured temporal responses in the transmission band contain several displacement wave packets (Figures 5c and S10), which originate from the reflections of the propagating acoustic wave at the ends of the waveguide. Meanwhile, at the stop band or bandgap frequencies, the mechanical displacement is below the noise floor of the system. Color map in Figure 5b is attained by measuring the time-of-flight (ToF) traces as a function of frequency across the first phonon band. Fringes rising from the multiple reflections can be resolved as well. Since the period of these fringes corresponds to the time that acoustic wave travels a round trip inside the waveguide ($\Delta t$), it can be utilized to deduce the experimental group velocity ($v_g$) of the acoustic wave packet, using the equation $\Delta t = 2L/v_g$ (where $L$ is the total length of the waveguide). Likewise, the separation between adjacent peaks ($f_{int}$) in the frequency-domain measurements and simulations is defined as $f_{int} = v_g/(2L)$. The round-trip traveling time $\Delta t$ calculated from the frequency-domain simulation is marked in Figure 5b as yellow squares and dash lines, which align perfectly with the measurement results. Frequency-dependent group velocities can be quantitatively extracted from Figures 5b and S11. At the frequencies adjacent to the stop band, the group velocities are one order smaller than the ones at the center frequencies of the transmission band. The group velocity is defined as the first-order derivative (slope) of the dispersion relation, as $v_g \equiv \partial \omega / \partial k$, where $k$ is the wavenumber and $\omega$ is the angular frequency. The slow phonon propagation near the edge of the transmission bands can thereby be interpreted to be resulted from the abruptly decreased slope of the dispersion curves around the band edges as shown in Figure 4c. Within the first transmission band, the group velocity culminates over 250 m/s at the frequencies around 22 MHz (Figure S11d). It can be seen from the ToF measurement (Figure 5c) that at such frequencies, the displacement amplitude of the acoustic wave packets is still detectable, even after circulating in the waveguide for 4 times, corresponding to a length over 1.2 mm (see Supporting Information and Figure S10 for detailed analysis). In this measurement, the acoustic wave undergoes both propagation loss and multi-reflection loss at the clamped ends. If only the propagation loss is considered, the effective propagation length would be even longer. We thus note, by increasing the number of cells, multi-reflection loss will be much reduced and propagation length can be increased, and the transmission attenuation at the bandgap frequencies of designed h-BN waveguides can be further enhanced (as predicted in Figure 2c).

The quasi-1D h-BN waveguide demonstrated here exhibits a prominently higher first band frequency (15–24 MHz) and wider bandgap opening (3 MHz), compared to its counterparts made from conventional materials, such as silicon nitride (Si$_3$N$_4$), with similar geometry design (Supporting Table S1) [35,36]. These favorable properties for high frequency (3–30 MHz) RF applications originate from the high Young's modulus and low mass density of h-BN, which bestow enhancement in mode-coupling nonlinearity of the resonators as well [29,30].

Moreover, the ultrahigh breaking strain limit of these 2D crystals allows them to be stretched much more significantly and offer wider operation bandwidth, which is unachievable by employing traditional materials with much lower breaking or fracturing strain limits [41]. According to our further simulations (Figure 6), the operation frequency of h-BN phononic waveguides can be easily expanded from high frequency (HF, 3–30 MHz) to very high frequency (VHF, 30–300 MHz) bands, by incorporating strain-engineered thin h-BN layers that follow the membrane model [39]. Especially when monolayer or odd-number layers (<10 layers) of h-BN thin flakes are employed, the piezoelectric effect may be expected to facilitate additional tunability and active devices may also be realized.

Besides the desirable features for classical RF signal routing and processing, the intriguing essential applications of the h-BN phononic waveguides lie in the quantum regime. Based on this prototypical demonstration of quasi-1D phononic waveguides, other functional devices, such as photonic-phononic crystals and optomechanical cavities [42], can be realized by further designing and engineering the periodic elastic energy potential of these vdW layered crystals. Embedding h-BN quantum emitters into such phononic waveguides creates an exceptional platform for exploring spin-qubit interaction and coherent information via photonic-phononic pathways [43,44,45].

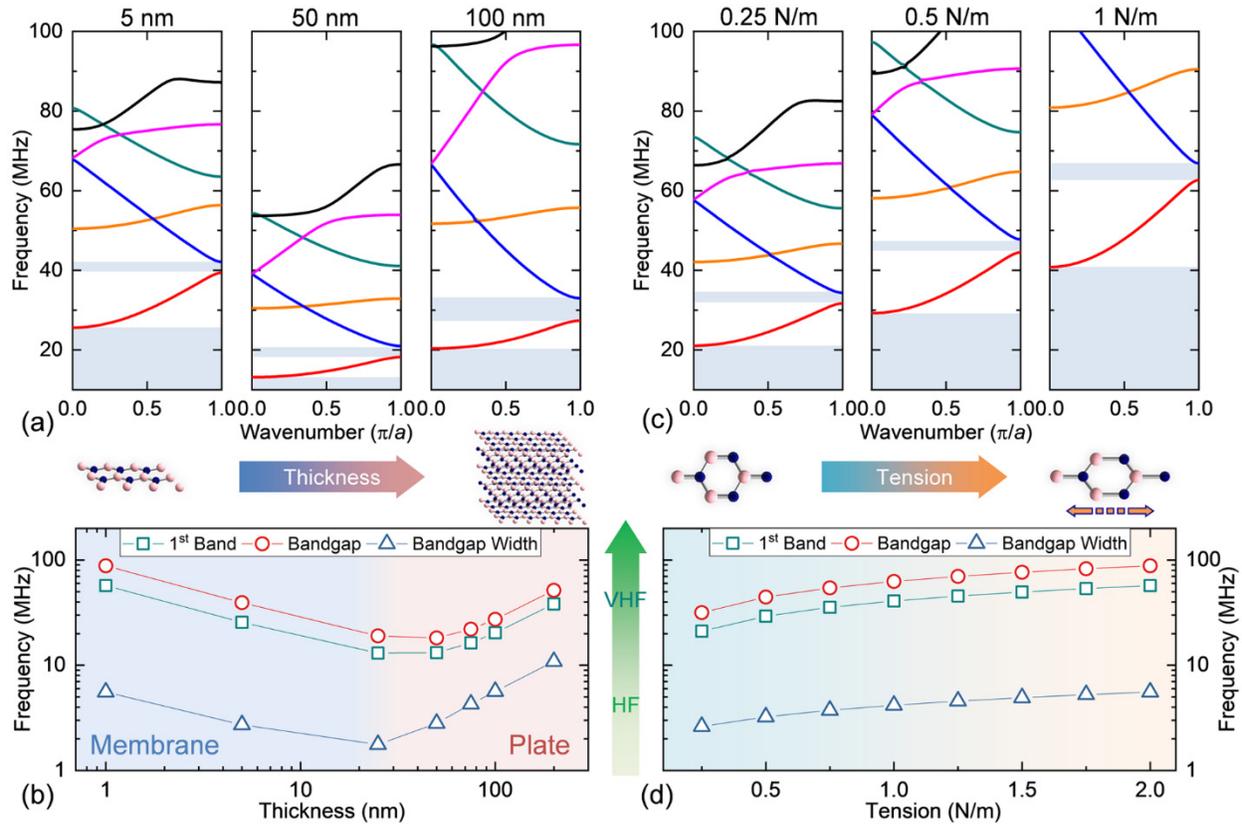

**Figure 6.** Tuning the phonon band structures with thickness and tension engineering. (a) Computed dispersion curves and (b) characteristic frequencies of phononic waveguides with the same unit-cell design (period *a*=6.785 μm, width *w*=10 μm, γ=2 N/m) but h-BN thickness varying from 1 nm to 200 nm. (c) Computed dispersion curves and (d) characteristic frequencies of phononic waveguides with the same unit-cell geometry (period *a*=6.785 μm, width *w*=10 μm, thickness *h*=1 nm), but tension level varying from γ=0.25 N/m to 2 N/m. Teal squares: Starting frequency of the first phonon band; Red circles: Starting frequency of the first bandgap; Blue triangles: Width of the first bandgap.

In conclusion, we have designed and developed the first phononic crystal waveguides in engineered h-BN nanomechanical structures. Taking advantage of the layered crystal structure, the challenges in nanofabrication of long-range-ordered phononic crystal lattices have been circumvented through a facile integration approach and elastic environment engineering. As-fabricated h-BN devices exhibit strong phonon dispersion relation with prominently higher first transmission band frequency and wider bandgap opening, compared to their counterparts made from conventional 3D crystalline materials with similar geometric designs. We have experimentally demonstrated that such h-BN waveguides are capable of supporting RF acoustic wave propagation over an effective length of 1.2 mm with a group velocity as high as 250 m/s at the first transmission band, while attenuating the transmission at the stop band and bandgap by over 30 dB. We envision by combining the unique piezoelectric properties of h-BN, such phononic structures can empower dynamically tunable devices for RF signal processing and open a new avenue towards building future integrated phononic and hybrid quantum circuitry. As the final remarks, the phononic crystal design and methodology developed in this work are directly transferable to other layered crystals beyond h-BN, such as the semimetal graphene and semiconducting transition metal dichalcogenides (TMDCs), and their heterostructures. The van der Waals integration of 2D materials also grants the applicability to produce such phononic devices onto any additive nanofabrication-compatible substrates (rigid, curved, and flexible), thereby empowering versatile functionalities.

## Supporting Information

Details of device design and fabrication, numerical simulations, measurement schemes, and data analysis can be found in Supporting Information.

## Corresponding Author

*Email: philip.feng@ufl.edu.

**Author Contributions**: Y.W. designed the devices. J.L. and Y.X. fabricated the devices. Y.W., J.L., and X.-Q.Z. carried out the device characterization with important technical support on apparatus and instrumentation from P.X.-L.F. Y.W., J.L., and P.X.-L.F. analyzed the data and wrote the manuscript. All authors reviewed and commented on the manuscript. P.X.-L.F. conceived the experiments and supervised the project.

## Acknowledgments

We are thankful for the support from the National Science Foundation (NSF) via EFRI ACQUIRE program (Grant EFMA 1641099), its Supplemental Funding through the Research Experience and Mentoring (REM) program, as well as the NSF CAREER Award (Grant ECCS 1454570).

*– Supporting Information –*

# Hexagonal Boron Nitride Phononic Crystal Waveguides

Yanan Wang,[†,‡] Jaesung Lee,[†,‡] Xu-Qian Zheng,[†,‡] Yong Xie,[†,l] and Philip X.-L. Feng[*,†,‡]

[†]*Department of Electrical Engineering & Computer Science, Case School of Engineering, Case Western Reserve University, Cleveland, Ohio 44106, United States*

[‡]*Department of Electrical & Computer Engineering, Herbert Wertheim College of Engineering, University of Florida, Gainesville, Florida 32611, United States*

[l]*School of Advanced Materials & Nanotechnology, Xidian University, Xi'an, Shaanxi 710071, China*

[*]Email: philip.feng@ufl.edu

In this Supporting Information for "Hexagonal Boron Nitride Phononic Crystal Waveguides", additional fabrication, simulations, and measurement details for the hexagonal boron nitride (h-BN) phononic devices are provided.

# I. Device Fabrication

In order to create long-range ordered suspended phononic structures from h-BN crystal, we develop a van der Waals integration approach. As discussed in the Main Text, traditional lithography-based techniques may induce photoresist (or electron-beam resist) residues that modify the surface and elastic properties of h-BN and hence degrade the device performance. A hybrid reactive ion etching (RIE) and electron beam-induced etching (EBIE) technique has been recently developed, but only been implemented to fabricate photonic and optomechanical devices with size limited to tens of micrometers and leaving rough sidewalls.[1,2]

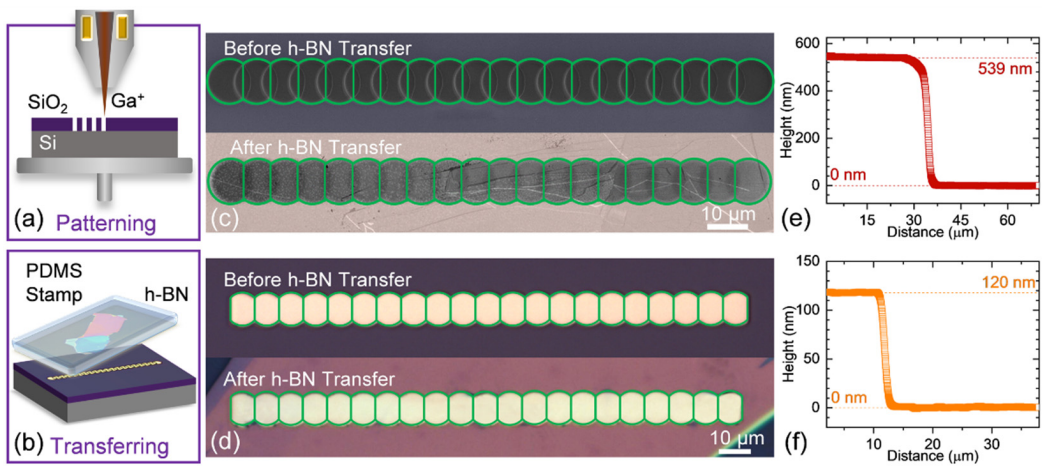

**Figure S1.** Fabrication of h-BN phononic waveguides. Schematic illustration of (a) substrate patterning via focused ion beam (FIB) etching and (b) polydimethylsiloxane (PDMS) assistant dry transfer steps. (c) Colored scanning electron microscopy images of a device with overlapped circular patterns before and after h-BN transfer. (d) Optical microscopy images of a device with simplified near oval-shaped unit cells before and after h-BN transfer. Both devices shown in (c)-(d) have a unit cell period of $a$ = 8.25 μm, width of $w$ = 12 μm. (e)-(f) Thickness measurement of the flakes shown in (c) & (d), respectively. Most measurements and simulations presented in the Main Text are conducted based on the geometry of the device shown in (d).

As illustrated in Figure S1a, to circumvent the aforementioned disadvantages, the periodic structures are defined on the commonly used oxidized silicon (290 nm $SiO_2$ on Si) supporting substrates, taking advantage of the well-established patterning and etching techniques with high spatial precision. Then, a suite of specially developed, completely dry exfoliation and transfer techniques are employed.[3] h-BN layers are mechanically isolated from a high-quality bulk h-BN (HQ Graphene) crystal by mechanical exfoliation and then pressed onto polydimethylsiloxane (PDMS) stamps. Large h-BN flakes with lateral size over 200 μm are identified under optical microscope and selectively transferred onto pre-patterned substrate locations with the aid of a micromanipulator to achieve mechanically suspended structures (Figure S1b). Figure panels S1c and S1d show the morphology of two devices before and after h-BN transfer. The thickness of h-BN flakes is measured by a stylus profilometer (KLA-Tencor P-6 stylus profilometer). The fabrication process is finished with optional high-temperature annealing at 850 ˚C to enhance the contact between h-BN layers and patterned substrates, and minimize the potential deterioration of device performance due to surface adsorbates.

## II. Waveguide Design and Numerical Simulations

### 1. Eigenfrequency and Frequency-Domain Simulations

The phononic waveguides are designed based on the finite element method (FEM) simulations (COMSOL Multiphysics). The phonon dispersion relation curves and propagation characteristics are simulated by applying the eigenfrequency and frequency-domain studies in the Solid Mechanics module, respectively. Mode shapes of the unit cell at different wavenumbers can also be attained from the eigenfrequency simulation and are helpful to interpret and assign the mode families to various branches of the dispersion curves. As shown in Figure S2, by analyzing the mode shapes of the unit cell at the wavenumber $k = 0$ and $k = \pi/a$ points, we identify mode crossing occurs for the phonon branches color-coded in blue and orange (Figure panels S2a and S2b), because the modes at the labeled points ③ and ⑥ share the same symmetry and belong to the same mode family.

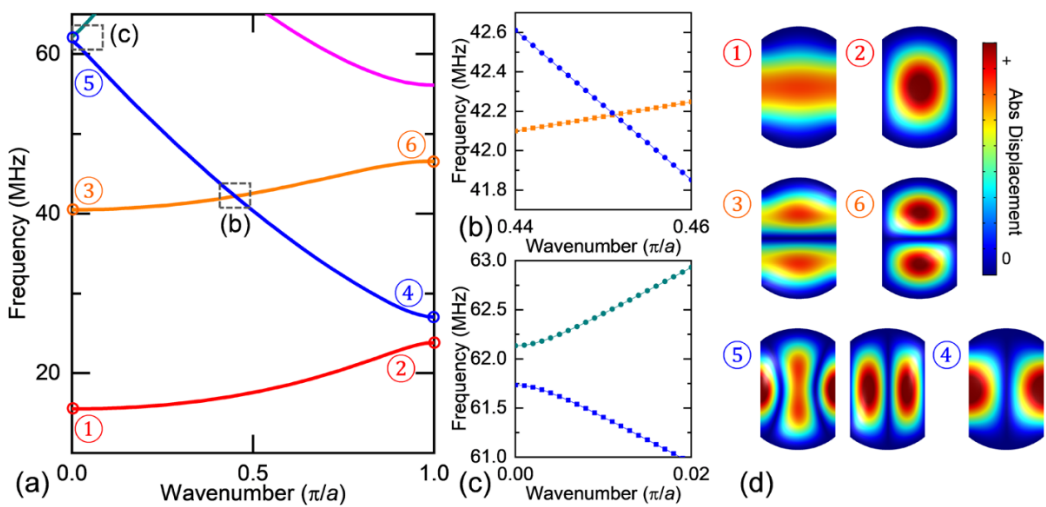

**Figure S2.** (a) Zoom-in plots of phonon dispersion curves shown in Figure 2b in the Main Text. (b) The mode crossing between the blue and orange curves and the anti-crossing between the blue and dark cyan curves are further zoomed in. (d) The unit-cell mode shapes for different branches at $k=0$ and $k=\pi/a$ points. Two possible mode shapes are presented for the mode anti-crossing at point ⑤.

### 2. Investigation of Coupling Strength

In order to obtain the desired response in the high frequency (HF, 3−30 MHz) band, the unit cell dimensions have been numerically investigated (Figure S3). It can be understood intuitively when the overlapping area is small (as in lattices with period $a = 10.5$ μm), the coupling strength is relatively weak and the resonators behave analogously to the individual ones, showing nondispersive frequency responses for odd-numbered modes. When the coupling strength overwhelms (as in lattices with period $a = 3$ μm), the resonators incline to vibrate as a whole with discrete frequency modes. With the aim of creating phononic structures with rich propagation characteristics, such as bandgap and mode crossing/anti-crossing, a unit cell design with moderate coupling strength, $a = 8.25$ μm, is chosen in this work.

### 3. Thickness and Tension Dependence

The influence of the thickness of h-BN layers on the frequency responses is illustrated in Figure 6 in the Main Text. Assuming the built-in tension of the h-BN layered crystal is uniform at $\gamma = 2$

N/m, the frequency bands downshift as the thickness increases for thinner flakes (thickness $h < 50$ nm). Whereas for thicker flakes ($h > 50$ nm), the frequency bands upshift as the thickness increases. Such thickness dependence is summarized in Figure 6b, and a clear transition from the membrane to plate model can be witnessed. Likewise, if the thickness of h-BN remains constant, the frequency response of the phononic waveguides can be tuned by varying the tension level inside the layered crystal. Taking h-BN with a thickness of 1 nm as an example, the bandgap frequency can be tuned by ~300% via varying the tension level from 0.25–2 N/m (Figure panels 6c and 6d), owing to the excellent structural properties of h-BN crystals. Moreover, it has been predicted that h-BN is a piezoelectric material when the thickness is down to the few-layer regime.[4] Combining the piezoelectric effects with these h-BN phononic waveguides can facilitate functional phononic devices with active control in the radio frequency (RF) band.

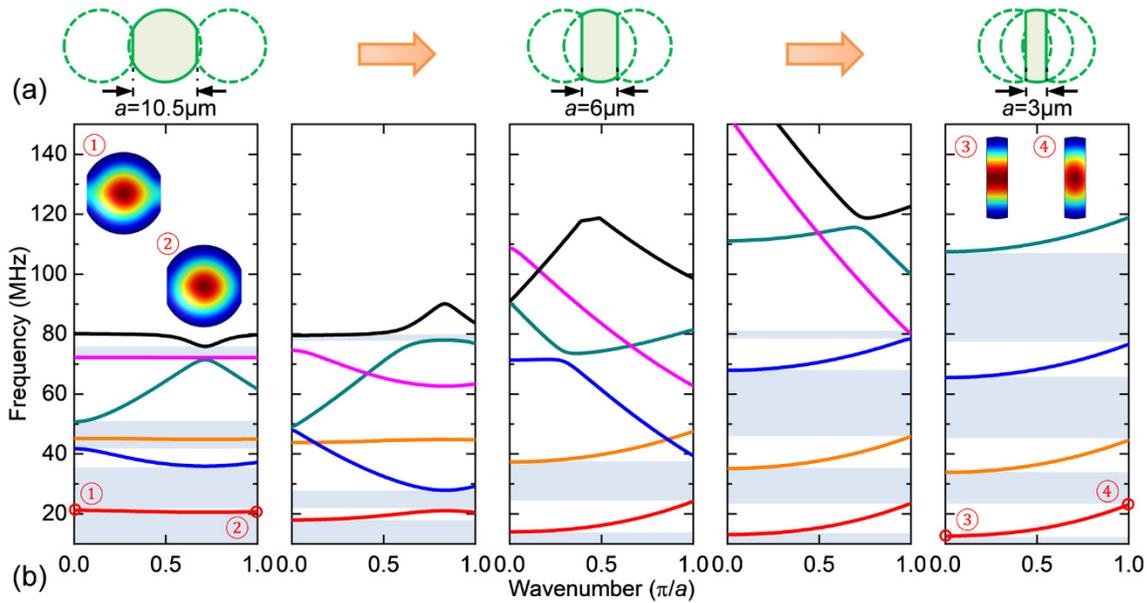

**Figure S3.** Engineering the phonon band structures via altering the coupling strength between adjacent cells. (a) Schematic illustration of controlling the coupling strength between neighboring resonators by varying the spatial overlapping. The effective unit cell highlighted in light green shade with the width kept as $w = 12$ μm and lattice period $a$ varying from 10.5 μm to 3 μm. (b) Resultant phonon dispersion curves and the mode shapes of unit cell (insets) for lattices with period $a = 10.5$ μm, 9 μm, 6 μm, 4.5 μm, and 3 μm (from left to right).

## III. Frequency-Domain Measurements

### 1. Measurement Scheme

The frequency-domain measurements of the as-fabricated h-BN phononic waveguides are performed in a customized ultrasensitive laser interferometry system (Figure S4a).[5] The h-BN phononic devices are secured inside a vacuum chamber with a quartz optical window, and the chamber is mounted on a motorized stage for precise position control. All the measurements in this work are performed at room temperature under a moderate vacuum of ~50 mTorr.

The interferometry system is equipped with an amplitude-modulated 405 nm blue laser as the photothermal driving beam and a continuous-wave 633 nm red laser as the detecting beam. The two laser beams are directed by a set of lenses, mirrors and beam splitters, and then focused onto the h-BN waveguide through a 50× long-working-distance objective. By adjusting the mirrors and beam splitters, the driving and detecting spots can be separated apart over 200 μm, thereby the frequency response along the waveguide can be resolved spatially. The reflected beam from the waveguide is collected via the same objective and detected by a photodetector (Newport, silicon detector 1801-FC) with a network analyzer. It is worth mentioning that the excitation laser spot is always parked at the supported h-BN region adjacent to the suspended waveguides, not on the suspended device area. As such, the excitation photons are more efficiently absorbed by the Si substrate underneath to generate localized photothermal actuation, considering the excitation photon energy (~3.06 eV) is far below the bandgap of h-BN (~5.9 eV). The detecting laser spot is parked at the center of the last cell; and its accurate position is optimized by monitoring the measured signal amplitude in the first phonon band.

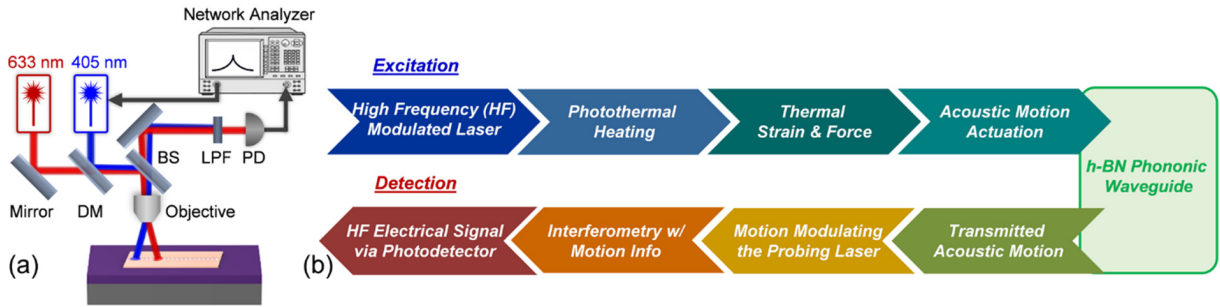

**Figure S4.** Frequency-domain measurement scheme. (a) Schematic of the customized laser interferometry system. The driving (405 nm) and detecting (633 nm) laser spots can be spatially separated up to ~200 μm under 50× magnifying objective. DM: Dichroic mirror; BS: Beam splitter; LPF: Long-pass filter; PD: Photodetector. (b) Signal transduction scheme.

## 2. Additional Measurement Results

Supporting Figure S5 presents two additional frequency-domain measurements from the device (period of $a$ = 8.25 μm, width of $w$ = 12 μm, thickness $h$ = 120 nm) discussed in the Main Text. Every spectrum consists of well-defined stop band, transmission band, and bandgap regions, and each peak in the transmission band can be fitted using the finite-$Q$ harmonic resonance function

$$v(f) = \Re_{\text{dis-vol}} \frac{F}{4\pi^2 m} \sqrt{\frac{1}{\left(f_0^2 - f^2\right)^2 + f_0^2 f^2 / Q^2}} + v_0, \quad (S1)$$

where $\Re_{\text{dis-vol}}$ is displacement-to-voltage responsivity (gain), $F$ is force, $m$ is effective mass, $f_0$ is center frequency of the peak, $Q$ is quality factor, and $v_0$ is the system background. The quality factors of the fitted resonance peaks range from 50 to 100. The variation in the displacement amplitude from multiple measurements may rise from the slightly different excitation and detection positions, depending on the relative distances from the node or antinode points (see Subsection III.3). While the peak center frequencies keep consistent among different times of measurement.

Devices with different geometric settings have also been fabricated and examined. Supporting Figure S6 shows a device with period of $a$ = 6.785 μm, width of $w$ = 10 μm and thickness of $h$ = 140 nm. Similar to the device investigated in the Main Text, the frequency spectrum of this waveguide shows clear stop band and bandgap features agreeing well with the numerical simulation. With a smaller unit cell, the first phonon band span over a higher frequency range from ~22–36 MHz, compared to the waveguide with $a = 8.25$ μm.

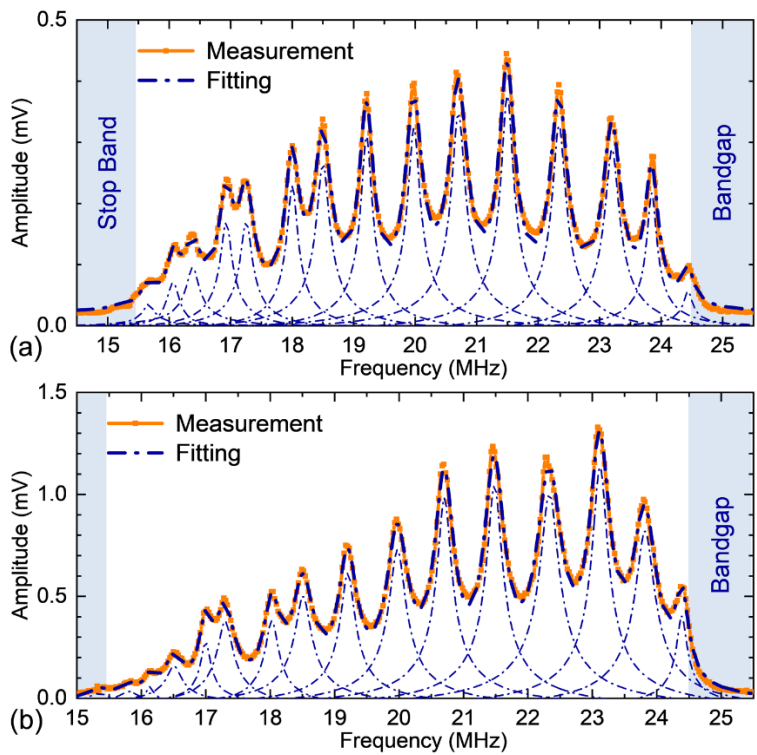

**Figure S5.** Measured frequency responses. Date in panels (a) and (b) are taken from two independent measurements with the same device (period of $a = 8.25$ μm, width of $w = 12$ μm, thickness $h = 120$ nm) shown in Figure S1d and the Main Text.

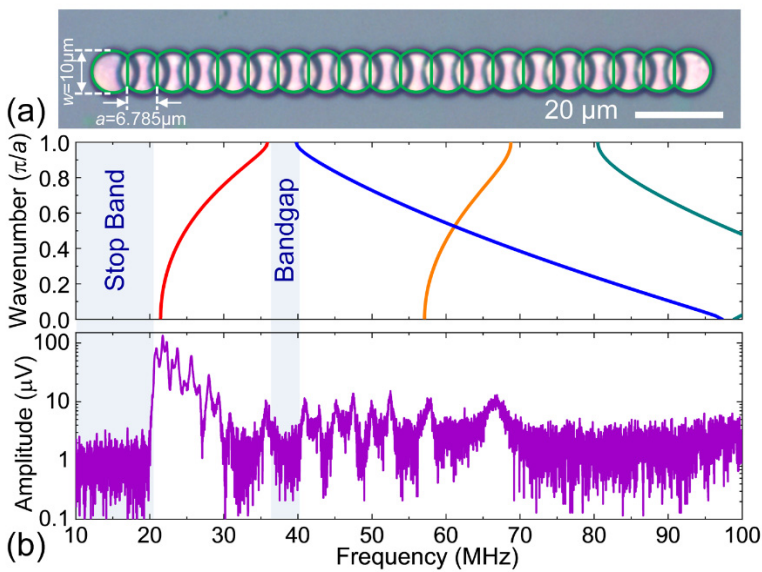

**Figure S6.** Example device with different geometric parameters. (a) Optical microscopy image of a device with period of $a = 6.785$ μm, width of $w = 10$ μm and thickness of $h = 140$ nm. (b) Computed dispersion curves (upper plot) and measured frequency response (lower plot) of the device shown in (a).

## 3. Mode Shape Dependency of Detection Efficiency

The influence of detecting laser spot position (within the cell being probed) on the measured displacement is further estimated via numerical simulation. Figure S7c illustrates the case in which the 633 nm laser is parked at the center of the cell. The detecting laser spot overlaps perfectly with the maximum displacement point (antinode) for the mode family of the first phonon band. This position, however, is on the nodal line for the modes of the second lowest phonon branch around $k = \pi/a$ point, at which the displacement is minimum. The dependency of computed displacement on the laser spot position is summarized as Figure S7d. The ratio of displacement amplitude between the second phonon band and the first phonon band is about $|u_④/u_②| \approx 0.151$ when the laser is focused at the center of the cell, which quantitatively matches the measured result of ~0.147 in Figure S7b. The ratio $|u_④/u_②|$ will gradually increase as the detecting laser spot moves along the white dashed line towards the edge of the cell, and can be intuitively interpreted from the mode shapes of these two phonon branches at $k = \pi/a$ point. Therefore, the actual or corrected amplitude of the $2^{nd}$ transmission band, had it not been compromised by the extrinsic detection efficiency drop (due to detection laser spot staying around nodal line), is about $|u_{④,\text{antinode}}/u_{④,\text{node}}| \approx 6.81$ times of what is shown in the data. The corrected frequency spectrum shown as dashed line in Figure 3b (Main Text) is attained by applying this correction factor.

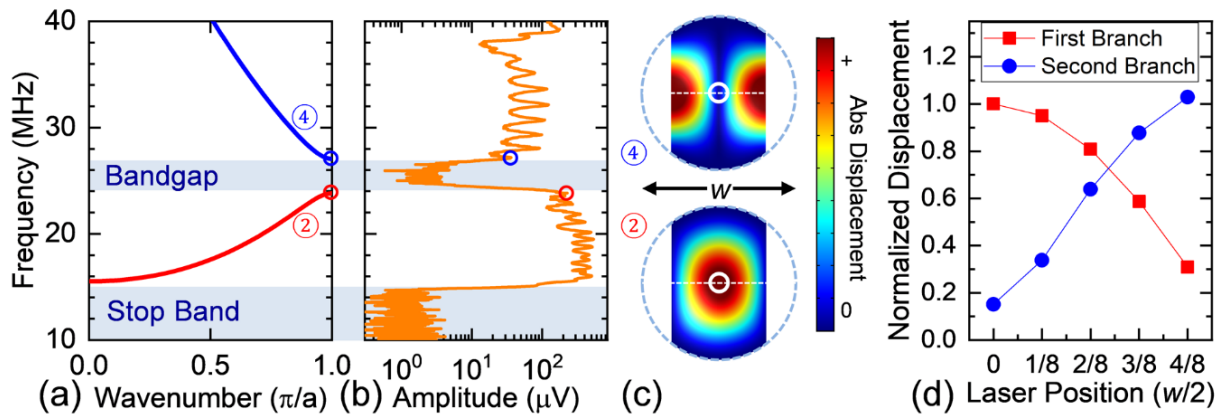

**Figure S7.** (a) Numerically computed dispersion characteristics. (b) Measured frequency spectrum in log scale. (c) Unit-cell views of mode shapes of the waveguide, for the first and second lowest phonon branches at $k = \pi/a$. The detecting laser spot position (with realistic spot size) is marked as a white circle with a diameter of ~1.5 μm. (d) Simulated displacement universally (for all modes) normalized to the antinode displacement of the mode at $k=\pi/a$ in the first phonon branch $|u_②|$, with laser position at the center. The laser position is plotted as the relative distance to the center of the cell in a unit of a half-width (radius = $w/2$) of the cell ($w = 12$ μm).

## 4. Cell Location Dependency of Frequency Spectra

By shifting the detecting 633 nm laser, we can measure the displacement of the waveguide at different cell locations (*i.e.*, counting number or index of cell, from left to right), and a typical set of frequency spectra are shown in Figure S8 from a device with $a = 6.785$ μm.

Because of the mode coupling amongst the cells, the frequency spectrum measured even from the first cell (probing laser parked at cell 1) appears to be collective mechanical modes, distinct from the discrete peaks of an individual resonator. However, the bandgap feature is barely resolvable (Figure S8a). As the detecting spot is moved further away from the excitation site (supported area closed to cell 1), the attenuation within the bandgap becomes more prominent, consistent with the frequency-domain simulation results. It is noticeable that the spectrum measured at cell 1 exhibits higher background level. This might result from more 'feedthrough' of the photothermal actuation effects experienced by cell 1, due to the very short distance from the motion excitation spot.

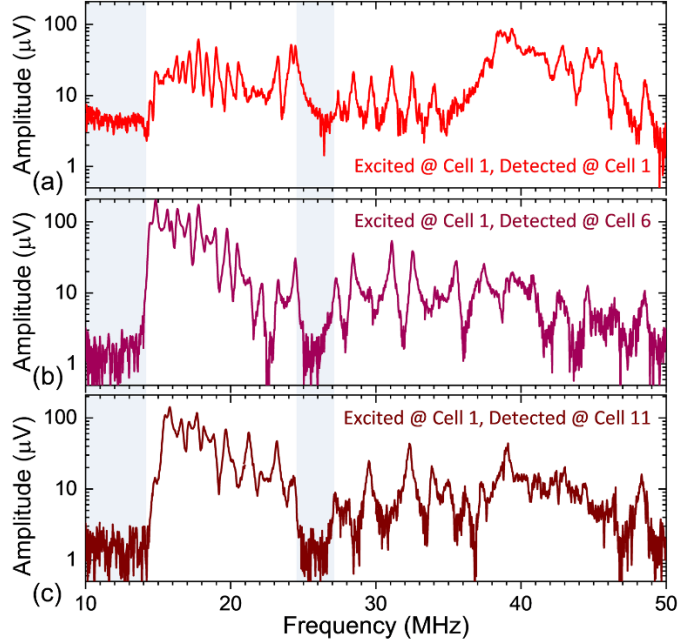

**Figure S8.** Frequency spectra of measured from a device (period of $a = 6.785$ μm, width of $w = 10$ μm, thickness of $h = 100$ nm, number of cells = 20) with fixed excitation laser position near unit cell 1, but with detecting (probing) laser at various cell locations along the waveguide, (a) unit cell 1, (b) unit cell 6, and (c) unit cell 11, respectively.

## 5. Displacement Estimation

The actual displacement of the waveguide can be estimated based on the responsivity of the laser interferometry system. First, we analytically calculate the light reflectance of the h-BN phononic crystal as follows:[6]

$$R_{\text{h-BN}}\left(\lambda\right) = \left| \frac{r_1 e^{i(\phi_1+\phi_2)} + r_2 e^{-i(\phi_1-\phi_2)} + r_3 e^{-i(\phi_1+\phi_2)} + r_1 r_2 r_3 e^{i(\phi_1-\phi_2)}}{e^{i(\phi_1+\phi_2)} + r_1 r_2 e^{-i(\phi_1-\phi_2)} + r_1 r_3 e^{-i(\phi_1+\phi_2)} + r_2 r_3 e^{i(\phi_1-\phi_2)}} \right|^2 , \tag{S2}$$

$$r_1\left(\lambda\right) = \frac{n_{\text{vac}}\left(\lambda\right) - n_{\text{h-BN}}\left(\lambda\right)}{n_{\text{vac}}\left(\lambda\right) + n_{\text{h-BN}}\left(\lambda\right)}, \ r_2\left(\lambda\right) = \frac{n_{\text{h-BN}}\left(\lambda\right) - n_{\text{vac}}\left(\lambda\right)}{n_{\text{h-BN}}\left(\lambda\right) + n_{\text{vac}}\left(\lambda\right)}, \ r_3\left(\lambda\right) = \frac{n_{\text{vac}}\left(\lambda\right) - n_{\text{Si}}\left(\lambda\right)}{n_{\text{vac}}\left(\lambda\right) + n_{\text{Si}}\left(\lambda\right)}, \tag{S3}$$

$$\phi_1 = 2\pi n_{\text{h-BN}} h / \lambda , \ \phi_2 = 2\pi n_{\text{vac}} d_{\text{vac}} / \lambda . \tag{S4}$$

In the above equations, $r_1$, $r_2$ and $r_3$ represent the reflection coefficients at the vacuum/h-BN, h-BN/vacuum, and vacuum/Si interfaces, and $n_{\text{vac}}$, $n_{\text{h-BN}}$, and $n_{\text{Si}}$, are the refractive indices of vacuum, h-BN, Si, respectively. The corresponding phase shifts are symbolized in $\phi_1$, $\phi_2$. Here $d_{\text{vac}} \approx 715$ nm is the depth of the vacuum gap between suspended h-BN and underneath Si surface, for devices in the Main Text.

Figure S9 shows calculated reflectance dependence on the depth of vacuum cavity or gap and the first derivative of curve gives the displacement-to-optical-reflectance responsivity, which is $\Re_{\text{dis-opt,PnC}} \approx \left| 0.136\% / nm \right|$ at 715 nm in this measurement. In our previous study of h-BN drumhead mechanical resonators, we obtain a displacement-to-optical-reflectance responsivity of $\Re_{\text{dis-opt,Res}} \approx \left| -0.846\% / nm \right|$.[7]

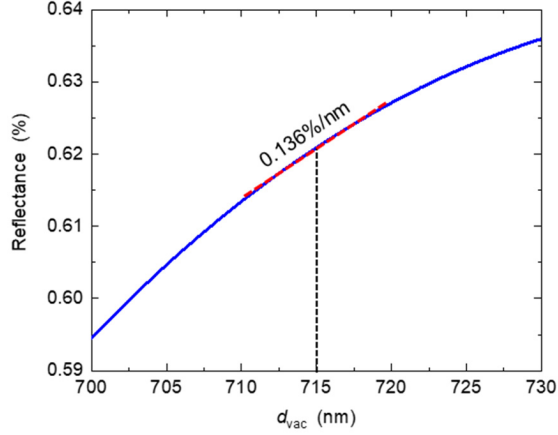

**Figure S9**. Calculated reflectance and responsivity based on the device shown in the Main Text.

To further quantitatively establish the conversion from voltage readout of network analyzer to the actual displacement, one method is to directly measure the undriven thermomechanical motion of the device as a calibration reference. However, due to the thickness of h-BN waveguide demonstrated in this work (>100 nm, in plate regime), the thermomechanical resonances of these devices are not readily resolvable. Nonetheless, we can use the displacement-to-voltage responsivity of $\Re_{\text{dis-vol,Res}} \approx 10.5$ nV/fm obtained from drumhead resonators with similar dimensions in Ref. 7 to do the calibration. Considering the same interferometry system is employed, $\Re_{\text{dis-vol}}$ should proportionally depend on $\Re_{\text{dis-opt}}$ between these measurements. Accordingly, the estimated displacement-to-voltage responsivity of the phononic waveguide device in the Main Text is

$$\Re_{\text{dis-vol,PnC}} \approx \left( \Re_{\text{dis-opt,PnC}} / \Re_{\text{dis-opt,Res}} \right) \times \Re_{\text{dis-vol,Res}} = 1.93 \text{nV/fm} . \tag{S5}$$

As shown in Figures 3 (Main Text) and Figure S5, the displacement readout of the phononic waveguides ranges from ~100 μV to ~1.5 mV. Therefore, the actual displacements of the devices are estimated to be on a sub-nanometer scale (~0.05 nm to 0.7 nm).

## 6. Transduction Analysis and Transmission Estimation

The signal transduction chain of our measurement system can be divided into three major components, including the upstream photothermal actuation, h-BN waveguide transmission, and the downstream motion detection modules, as illustrated in Figure 4a (Main Text). A detailed transduction scheme involving each signal conversion step is analyzed and described as follows.

The input voltage signal $v_{\text{in}} \cos(\omega t)$ from the network analyzer to the photothermal actuation module will be converted in an optical domain signal based on the setting of the controller of the amplitude modulated 405 nm laser. In our measurement, the input from the network analyzer (RF output driving power at $P_{\text{dr}}$, measured in dBm) is $v_{\text{in}} = 200$ mV, and a modulation depth of the optical signal is set to be 100%. Then the signal transduction from electrical to the optical domain can be written as

$$v_{in} \cos(\omega t) \rightarrow P_{405nm} \left[ 1 + \cos(\omega t) \right]. \tag{S6}$$

Here, $P_{405nm}$ is the power of the 405 nm laser and measured as 3.3 mW. The modulated laser light is then significantly attenuated in the optical circuit, focused at the actuation spot on the device chip, absorbed by the Si substrate, consequently the temperature of the Si will be periodically elevated based on the modulation frequency.

$$\Delta T(t) \propto a_1 \eta c_1 P_{405nm} \cos(\omega t), \tag{S7}$$

where $a_1$ is the absorbance of Si at 405 nm wavelength, $\eta$ is a non-radiative recombination rate, $c_1$ is a constant determined by the device structure (such as geometry). Considering low thermal conductivity of $SiO_2$ ($\kappa_{SiO2} \approx 1.4 \times 10^{-3}$ W cm$^{-1}$ °C$^{-1}$) and sizeable thermal contact resistance between the substrate and h-BN layers, we assume the heat transfer from the Si substrate to the suspended h-BN is comparatively negligible.

The temperature variation of the Si substrate is then converted to the mechanical deformation through the thermal expansion and the displacement of Si, $u_{Si}$, which is

$$u_{Si}(T) \propto \alpha_{Si} \Delta T(T), \tag{S8}$$

where $\alpha_{Si}$ is thermal expansion coefficient of Si. The mechanical motion of Si $u_{si}$ then translates into the mechanical motion of the h-BN $u_{h\text{-}BN}$, which propagates in the waveguide back and forth. At a standing wave condition (*i.e.*, constructive interference of the mechanical wave in h-BN), the displacement of h-BN will be amplified by $Q$ factor of the waveguide:

$$u_{h\text{-}BN}(T) \propto Q u_{Si}(T). \tag{S9}$$

The mechanical motion of the h-BN is then converted to the optical domain based on the laser interferometry condition between mechanical wave and the 633nm laser light. The signal in optical domain is then transduced to electrical voltage $v_{out}$ via the photodetector and read out by a network analyzer.

Because of the many components involved in the motion detection scheme above and the complexity of the energy conversion in such components, it is impractical to accurately attain this transduction gain or ratio purely analytically. However, the total transmission of this measurement scheme can be simplified as the sum of the device and the system conversion gain/loss from actuation and detection modules, which can be directly obtained from the measurement by using the equation $T_{out\text{-}in} = 20 \log(|v_{out}/v_{in}|)$ as shown in Figure 4e (Main Text). In order to estimate the intrinsic transmission of phononic waveguide (transmission due to the acoustic wave propagation) solely, we need to subtract the extrinsic contributions from the total transmission. Following a similar strategy for displacement estimation, we employ a well-investigated h-BN drumhead resonator as reference and calibrate the transduction gain or conversion loss of the system.[7] The h-BN drumhead resonator shares similar diameter as the unit cell of the phononic waveguide and is measured under similar excitation and detection conditions in the same laser interferometry system. Therefore, we assume the system transmission can be canceled out by comparing the total transmission of these two devices and compensating the differences in the optical interferometry responsivity ($\Re_{dis\text{-}opt}$) (as discussed in Subsection III.5). For the phononic waveguide, the device transmission ($T_{PnC}$) contains both localized vibration term ($T_{Loc}$) and propagation term ($T_{N\text{-}1}$). For the drumhead resonator, the device transmission is only related to the localized vibrations. By assuming the localized vibration term is directly proportional to the device quality factor ($Q$), the propagation term of the waveguide can be written as

$$T_{\text{N-1}} = T_{\text{out-in,PnC}} - \left(T_{\text{out-in,Res}} + 20\log\left(Q_{\text{PnC}} / Q_{\text{Res}}\right) + 20\log\left(\Re_{\text{dis-opt,PnC}} / \Re_{\text{dis-opt,Res}}\right)\right) \text{ [dB]}. \quad \text{(S10)}$$

The total transmission of the drumhead resonator at a frequency of 19.4 MHz, lying within the first pass band of the waveguide shown in the Main Text (15–24 MHz), is measured as[7]

$$T_{\text{out-in,Res}} = 20\log\left(\left|v_{\text{out,Res}} / v_{\text{in,Res}}\right|\right) = 20\log(1.7\text{mV}/200\text{mV}) \approx -41.4\text{dB}. \quad \text{(S11)}$$

With substituting parameters of $Q_{\text{PnC}} \approx 100$, $Q_{\text{Res}} \approx 329$, $\Re_{\text{dis-opt,PnC}} \approx \left|0.136\% / \text{nm}\right|$, and $\Re_{\text{dis-opt,Res}} \approx \left|-0.846\% / \text{nm}\right|$, the intrinsic transmission of the phononic waveguide shown in the Main Text can be estimated as $T_{\text{N-1}} \approx T_{\text{out-in,PnC}} + 67.6$ dB (disregarding the frequency-dependent system response). Figure 4f in the Main Text is attained by applying the above calibration relationship to the measured data set in Figure 4e.

## IV. Time-Domain Measurements

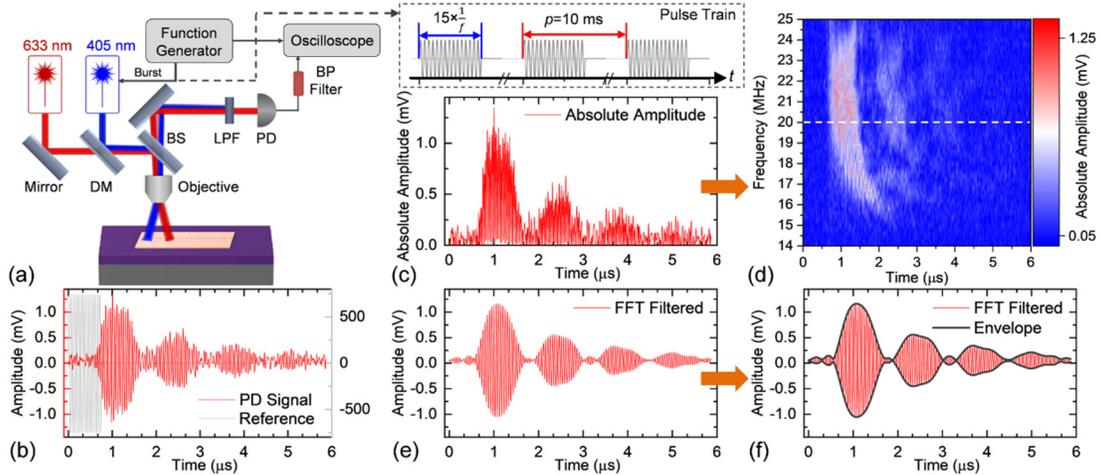

**Figure S10**. Time-domain measurement scheme. (a) Schematic of the modified system with function generator providing the burst signal and oscilloscope reading out the PD signals. DM: Dichroic mirror; BS: Beam splitter; LPF: Long-pass filter; PD: Photodetector; BP: Band-pass. The call-out shows the pulse train of the burst (reference) signal from function generator, with duration of $15 \times f^{-1}$ and period $p = 10$ ms. (b) Typical time-of-flight (ToF) trace at the frequency of 20 MHz. (c)-(d) Data processing to attain the color map of time-domain responses at different frequencies. The white dash line in (d) marks the frequency of 20 MHz. (e)-(f) Data processing to extract wave envelopes (black curves in (f)) from the ToF trace.

To monitor mechanical wave propagation in the time domain, we use the measurement configuration shown in Figure S10a. The modulation frequency of the 405 nm laser is controlled by a function generator. We introduce a single frequency sinusoidal signal of 15 cycles by setting the RF burst mode of the function generator. The RF burst signal gradually increases the displacement of the device, and then the generated wave packet propagates through the phononic waveguide until it relaxes to its thermal noise limit. The displacement of the mechanical wave packet is probed at the end of the waveguide using the 633 nm red laser and converted to the electrical signal using a photodetector. The excitation signal from the function generator and the displacement from the waveguide are simultaneously recorded using an oscilloscope. A set of results are averaged over 512 times by synchronizing the time domain signals using the trigger from the function generator to the oscilloscope. The period of the burst signal is set as 10 ms so as to monitor the complete relaxation process and evolution of the wave packet while it propagates back and forth inside the waveguide.

Typical time-of-flight (ToF) trace is exemplified in Figure S10b as multiple wave packets with gradually decreasing amplitudes. The wave pulse profile becomes slightly distorted after several round trips because of the multiple interferences. In order to attain the color map shown as Figure 5b in the Main Text, the measured amplitude trace at each frequency is converted to the absolute amplitude plot (Figure S10c) and then stacked together. The white-red fringes in the color plot correspond to the wave packets in the ToF traces. Small group velocities can be witnessed in the color map, especially at the frequencies close to the lower edge of the transmission band.

As discussed in the Main Text, the separation between two neighboring wave packets is equivalent to the time that acoustic wave travels a round trip inside the waveguide $\Delta t = 2L/v_{\mathrm{g}}$ (where $v_{\mathrm{g}}$ is the group velocity, $L$ is the total length of the waveguide). The wave envelopes extracted from the ToF traces can be utilized to estimate the group velocity. For example, the group velocity derived from Figure S10f is ~250 m/s for the acoustic wave frequency of 20 MHz, matching the value calculated based on the frequency-domain simulation (Figure S11d). The total loss, including both propagation and reflection loss, can also be approximated from the ToF traces. The amplitude decay between the first two neighboring packets at 20 MHz is calculated as ~51% in voltage and ~76% in energy for a round trip of ~350 μm, corresponding to ~17 dB/mm.

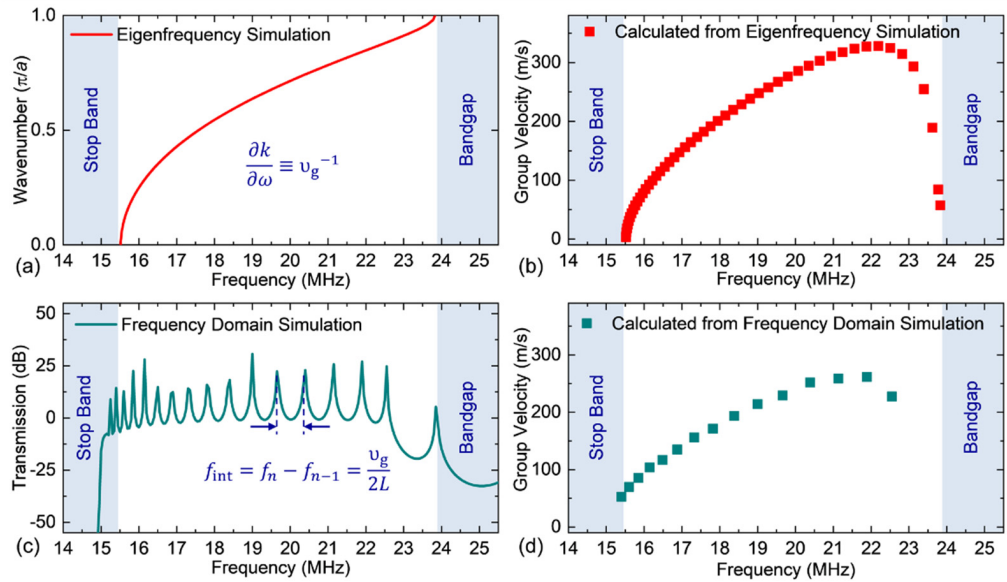

**Figure S11**. Analysis and calculation of the dispersive group velocities. (a)-(b) Calculation from the first-order derivative of the simulated phonon energy dispersion curve. (c)-(d) Calculation from the separation between the neighboring peaks in the frequency-domain simulated spectrum.

## V. Comparison with State-of-the-Art Phononic Waveguides

To summarize this work, we also make a comparison between h-BN phononic waveguides demonstrated here with some state-of-the-art thin-film phononic waveguides made in other materials. It is noticeable that our demonstration of h-BN phononic waveguides possess higher first band frequency and larger bandgap width compared to their counterparts made in conventional materials, with similar geometry design.[9,10] As discussed in the Main Text, these high-frequency characteristics originate from the high Young's modulus (theoretical value of $E_Y$ =780 GPa,[8] fitted value of $E_Y$ =370 GPa from experiments in this work, consistent with $E_Y$ values determined in earlier experiments[7]) and low mass density (2,100 kg/m³) of the h-BN crystals. Such unique combination of mechanical properties also renders an ultrahigh speed of sound (>$1.3\times10^4$ m/s) and promises for enhancement in mode-coupling nonlinearity of the resonators. Moreover, the ultrahigh breaking strain limit (>20%) of these 2D crystals allows them to be stretched significantly and offer unusually wide operation bandwidth and tunability. As predicted in Figure 6 (Main Text), the operation frequency of h-BN phononic waveguides can be easily expanded from high frequency (HF, 3–30 MHz) to very high frequency (VHF, 30–300 MHz) bands, by incorporating thin h-BN layers which follow the membrane model. When monolayer or odd-number layers (<10 layers) of h-BN thin flakes are employed, the piezoelectric effect will facilitate dynamic tunability. Taking monolayer h-BN flake as an example, up to ~300% tuning in the starting frequency of the first bandgap can be attained via varying the tension level from 0.25N/m to 2 N/m (Figure 6c and 6d in Main Text). This level of tunability is unachievable by employing traditional crystalline materials, such as Si and GaAs, with much lower breaking or fracturing strain limits. Therefore, we envisage h-BN to be an excellent and promising material platform for implementing passive and active phononic devices, towards on-chip integrated circuits for both classical and quantum information processing.

**Table S1.** Comparison between the h-BN phononic waveguides demonstrated in this work and state-of-the-art thin-film phononic waveguides in other materials.

| | Material | Device Dimensions | # of Unit Cells | First Band Frequency [MHz] | Bandgap Width [MHz] |
|---|---|---|---|---|---|
| This Work | h-BN | $w$=12 μm, $a$=8.25 μm, $h$=120 nm | 21 | 15–24 MHz | ~3 MHz |
| | | $w$=10 μm, $a$=6.875 μm, $h$=140 nm | 20 | 22–36 MHz | ~4 MHz |
| | | $w$=12 μm, $a$=8.25 μm, $h$=540 nm | 20 | 35–50 MHz | ~10 MHz |
| Ref. [9] | SiN$_x$ | $w$=20 μm, $a$=7 μm, $h$~100 nm | 30, 60, 90, 120 | 12–17 MHz | ~2.8 MHz |
| Ref. [10] | GaAs/AlGaAs Heterostructure | $w$=22 μm, $a$=8 μm, $h$~200 nm | 125 | 3.5–7.5 MHz | ~0.9 MHz |